\documentstyle[12pt]{article}
\newlength{\dinwidth}
\newlength{\dinmargin}
\begin{document}
\begin{center}
{\bf \huge Slow proton production in
semi-inclusive deep inelastic scattering and the pion cloud in the nucleon \\}

\vspace{1cm}

{\bf Antoni Szczurek}
 \medskip\\
{\small \sl
Institute of Nuclear Physics, PL-31-342 Cracow, Poland}

\vspace{1cm}

{\bf Gerard D. Bosveld} and {\bf Alex E.L. Dieperink}
 \medskip\\
{\small \sl
Kernfysisch Versneller Instituut, NL-9747 AA Groningen, The Netherlands}

\vspace{1cm}

{\bf \LARGE A b s t r a c t \bigskip\\}
\end{center}

The  semi-inclusive cross section for producing slow protons
in charged current deep inelastic
(anti-) neutrino scattering on
protons and neutrons is calculated as a function of the Bjorken $x$ and
the proton momentum. The standard hadronization models
based upon the colour neutralization mechanism appear to
underestimate the rate of slow proton production on hydrogen.
The presence of the virtual mesons (pions) in the nucleon leads to
an additional mechanism for proton production, referred to as spectator
process.
 It is found that at low proton momenta both
mechanisms compete, whereas  the spectator mechanism dominates at very
small momenta, while the color neutralization mechanism dominates
at momenta larger
than 1-2 \, GeV$/c$. The results of the calculations are compared with
the CERN bubble chamber (BEBC)  data.
The spectator model predicts a sharp increase of the semi-inclusive
cross section at small $x$ due to the sea quarks in virtual mesons.

\vskip 2cm

PACS numbers: 25.30.Pt, 13.85.Hi


\newpage


\section{ Introduction}

Inclusive deep inelastic lepton scattering off nucleons is a well
established tool for investigating the parton distributions
\cite{TKF91}.
Hadrons, which are not measured in inclusive experiments, are
produced mainly in the colour neutralization process; the string
models being the state of art.
It is expected that the measurement of final state hadrons will lead to a
better understanding of the mechanism of hadron production as it contains
information about hadronization process.

In semi-inclusive deep inelastic reaction (SIDIS) one observes one or
more hadrons in coincidence with the scattered lepton.
In the simplest approximation the semi-inclusive cross section
is assumed to factorize into an inclusive cross section (proportional
to the deep-inelastic structure functions) and a fragmentation
function \cite{FF78,BFM82}.

In the past the study of semi-inclusive deep-inelastic lepton scattering
has been restricted mainly to the detection of high energy hadrons,
which originate from the fragmentation of the leading (struck) quark.
The production of slow protons in coincidence with muons has been
studied in the (anti)neutrino induced charged current reactions
at CERN \cite{A83,G89}.
In a more recent analysis of the CERN experiment WA21
\cite{G89} the fraction of slow
($p <$ 0.6 GeV$/c$) protons was determined. In Ref.\cite{MTN92}
the rate from Ref.\cite{G89} was reproduced with a constant fragmentation
function, assuming that only protons (no other
baryons) are produced. The rate of slow
proton production strongly depends on the behaviour of the fragmentation
function at the kinematical limit, i.e. the light cone momentum fraction
$z=p^-/m$ close to 1. The counting
rules \cite{BB74} suggest  for the diquark fragmentation function rather
a linear, $D_{2q}(z) = A(1-z),$
than a constant dependence on $z.$
Furthermore, other baryons (neutrons, isobars, etc.) can be produced
which leads to a suppression of the relative proton yield.

In order to illustrate
this point we show in Table 1 the relative yield of protons and
neutrons including both direct and sequential (intermediate $\Delta$)
production obtained from a more realistic, yet simple,
quark-diquark model \cite{BDT93}. For comparison we give the number
of protons/neutrons per event obtained from the Lund string model
\cite{I92}.
Both arguments discussed above suggest that the agreement with
experimental rate in Ref.\cite{MTN92} has to be reconsidered.
It will be discussed later that a more realistic calculation leads to a
deficit of slow protons compared to the experimental rate \cite{G89}.
The violation of the Gottfried sum rule  observed by NMC \cite{A91}
strongly suggests the flavour asymmetry of the sea of
light quarks in the proton. This can be naturally accounted for
by the presence of the pion (meson) cloud in the nucleon
\cite{HM90} -- \cite{SSG93}.
The pion cloud model
has also been used successfully in inclusive DIS to explain the antiquark
distributions \cite{SSG93,HSS94}.

The presence of the meson cloud may lead to an
additional mechanism for the slow proton production in deep inelastic lepton
scattering \cite{LS78,LPR79,KDS92,BDS92}. For instance,
in the (anti)neutrino induced reaction the virtual boson $W$ smashes
the virtual pion (meson) into debris and the nucleon (baryon) is
produced as a spectator of the reaction (see Fig.1). The presence of
the $\pi\Delta$ Fock component in the nucleon leads to the production
of a spectator $\Delta$ which decays into a pion and nucleon.

Up to now no fully consistent treatment of both the standard
fragmentation mechanism and the meson exchange mechanism exists.
We note that it is inconsistent to add the cross sections for the two
processes. In the present paper we discuss in a simple hybrid model how
the two mechanisms can be taken into account in a consistent
manner which ensures baryon number conservation.
It is also not fully consistent to replace the
four-quark fragmentation by the meson exchange mechanism due to
the following two reasons. First of all, as discussed in
Ref.\cite{SS93}, the virtual pions (mesons) give rise both
to the sea and valence quark distributions. Secondly
\cite{SSG93,HNSS94} the sea generated by mesons constitute only
a fraction of the total sea.

It is the aim of this paper to quantitatively predict the contribution
of the meson exchange mechanism to the slow proton production rate
in \hbox{(anti-)}neutrino charged current deep inelastic scattering from
nucleons.
In addition the momentum distribution of protons is calculated and
compared with existing experimental data \cite{A83}.
Finally the $x$-dependence of the coincidence cross section for
the experimental momentum bins is calculated and compared with experimental
data from Ref.\cite{G89}.

\newpage


\section{Deep inelastic (anti-)neutrino scattering}

\subsection{Inclusive scattering}
In leading order electroweak theory the inclusive, differential,
charged current (cc) cross section for the
neutrino/antineutrino scattering ($h(\ell,\ell')X$) can be expressed
in terms of the structure functions $F_1, F_2, F_3 $
\begin{equation}
\frac{d^3\sigma_{\rm cc}^{\ell}} {dx dy dQ^2} =
\frac{G_{F}^2}  {\pi (1 + \frac{Q^2}{  M_{W}^{2}}) }
[
y^2 x F_{1}(x,Q^2) +
 (1-y)F_{2}(x,Q^2) \pm
  (y - \frac{y^2}{2}) x F_{3}(x,Q^2)
].
\label{ewcs}
\end{equation}
Here $G_{F} = \pi / (\sqrt{2} \sin^2\theta_{W} M_{W}^{2})$ is the Fermi
coupling constant, $M_{W}$ is the $W$-boson mass and
$y=(E_{\ell'}-E_{\ell})/E_{\ell}$.
Assuming the Callan-Gross relation, $2xF_{1} = F_{2}$ and
the standard parton model the $y$ integrated cross section is given by
\begin{equation}
\frac{d^2\sigma_{\rm cc}^{\ell}} {dx dQ^2} =
\frac{ G_{F}^2 } { \pi (1 + Q^2/M_{W}^{2}) } x
{\cal F}_{\rm cc}^{\ell}(x,Q^2) \, ,
\end{equation}
where ${\cal F}_{\rm cc}^{\ell}(x,Q^2)$
can be expressed in terms of the quark distributions
\begin{eqnarray}
{\cal F}_{\rm cc}^{\nu}(x,Q^2)
& = &
d(x,Q^2) + \overline{u}(x,Q^2) +
{1 \over 2}(-y + {y^2 \over 2}) [ d(x,Q^2) + 3\overline{u}(x,Q^2) ] ,
\nonumber \\
{\cal F}_{\rm cc}^{\overline{\nu}}(x,Q^2)
& = &
\overline{d}(x,Q^2) + u(x,Q^2) +
{1 \over 2}(-y + {y^2 \over 2}) [ \overline{d}(x,Q^2) + 3u(x,Q^2) ]
\end{eqnarray}
for neutrino and antineutrino induced reactions, respectively.
In the expressions above the distributions of heavier quarks
 have been neglected.
In the scaling limit and transferred four-momenta $Q^2 \ll M_{W}^2$
the $Q^2$-integrated cross section is
\begin{equation}
\frac{d\sigma_{\rm cc}^{\ell}} {dx} = \frac{ G_{F}^{2} m_N E_{\rm beam} }{\pi}
{\cal G}_{\rm cc}^{\ell}(x),
\label{in_on_shell}
\end{equation}
where
\begin{eqnarray}
{\cal G}_{\rm cc}^{\nu} (x) = 2 x [ d(x) + \overline{u}(x)/3 ] ,
\nonumber \\
{\cal G}_{\rm cc}^{\overline{\nu}} (x) = 2 x [ \overline{d}(x) + u(x)/3 ] .
\end{eqnarray}
This formula will be useful in the further analysis.
\subsection{Semi-inclusive scattering: fragmentation approach}
Semi-inclusive charged current (anti)neutrino reactions, where
a slow proton is measured in coincidence with the muon, are of special
interest.
 The $Q^2$-integrated cross section for the production of a baryon
$B$ (nucleon, delta) can be written as
\begin{eqnarray} \lefteqn{
\frac { d\sigma[N(\ell,\ell'B)X] } {dx dz dp_{\perp}^2} =
\frac { G^2 m_{N} E_{\rm beam} } {\pi} } \nonumber \\
&& \sum_f \bigl[ V_{f}^{N(\ell,\ell')}(x) D_{2q}^{B/f}(\tilde{z},p_{T}^2) +
        S_{f}^{N(\ell,\ell')}(x) D_{4q}^{B/f}(\tilde{z},p_{T}^2) \bigr]/(1-x).
\label{semiinclcs}
\end{eqnarray}
Here $V_{f}^{N(\ell,\ell')}(x)$ and $S_{f}^{N(\ell,\ell')}(x)$ are
functions proportional to the valence and sea quark distributions,
respectively, with coefficients being reaction and flavour ($f$) dependent,
 $D_{\rm debris}^{p}(\tilde{z},p_{T}^{2})$ is the target debris
$\rightarrow$ proton fragmentation function, with $z = p^{-}/m_{N},
\ \tilde{z}=z/(1-x),$ and $p_{T}$ is
the transverse momentum of the emitted proton (baryon).
Usually a factorized form of the fragmentation function is assumed
\begin{equation}
D_{\rm debris}^{p}(\tilde{z},p_{T}^{2}) = D(\tilde{z}) \Phi(p_{T}^{2}) .
\label{phi}
\end{equation}
Then, neglecting a small production of the $B\overline{B}$ pairs,
the baryon number conservation requires for each
flavour the following sum rules (assuming only one baryon is produced
in the target fragmentation region)
\begin{eqnarray}
\sum_{B} \int D_{2q}^{B/f}(z) \, dz = 1 , \nonumber\\
\sum_{B} \int D_{4q}^{B/f}(z) \, dz = 1 ,
\label{BNC}
\end{eqnarray}
separately for diquark, $D_{2q}(z),$ and four-quark, $D_{4q}(z),$
fragmentation functions.

In more advanced string models \cite{AGIS83,W84,AGN87,W89,WK90}
the semi-inclusive cross sections can be calculated with the help
of Monte Carlo methods \cite{S86,S87,W90,I92}.


\subsection{Pionic contribution to the semi-inclusive
cross section}


In the simplest models the nucleon is treated as a system of
three quarks. The pion-nucleon interaction leads naturally to
an admixture of a $\pi N$ Fock component in the physical
nucleon. In the simplest approximation, the Fock state decomposition
of the hybrid proton (neutron) reads
\begin{eqnarray}
|p\rangle&=&\cos\theta|(p_{0})\rangle+\sin\theta
      \left[\sin\phi\left(\sqrt{2/3}|\pi^+ n\rangle
        -\sqrt{1/3}|\pi^0 p\rangle\right)\right.       \nonumber\\
&&\left.+\cos\phi\left(\sqrt{1/2}|\pi^-\Delta^{++}\rangle-
\sqrt{1/3}|\pi^0\Delta^{+}\rangle
+\sqrt{1/6}|\pi^+\Delta^{0}\rangle\right)\right] \;\ ,
\label{WFp}
\end{eqnarray}
\begin{eqnarray}
|n\rangle&=&\cos\theta|(n_{0})\rangle+\sin\theta\left[\sin\phi
      \left(\sqrt{1/3}|\pi^0 n\rangle
      -\sqrt{2/3}|\pi^- p\rangle\right )\right.  \nonumber\\
&&\left.+\cos\phi\left(\sqrt{1/2}|\pi^+\Delta^{-}\rangle-
\sqrt{1/3}|\pi^0\Delta^{0}\rangle
+\sqrt{1/6}|\pi^-\Delta^{+}\rangle\right)\right] \;\ ,
\label{WFn}
\end{eqnarray}
where only $\pi N$ and $\pi \Delta$ Fock states are taken into account.

The two mixing angles are related to the number of pions
$(n_{\pi}=n_{\pi N}+n_{\pi \Delta})$ in the pionic cloud
\begin{eqnarray}
\sin^2\theta=n_{\pi}/(1+n_{\pi})\nonumber\\
\sin^2\phi =n_{\pi N}/n_{\pi} \;\ . \nonumber
\end{eqnarray}
The Fock expansion in Eqs. (\ref{WFp}, \ref{WFn}) leads to
the processes which can be classified as spectator mechanism
(see Fig.1), core fragmentation, and intermediate nucleon
(and intermediate $\Delta$) fragmentation with the pion produced
as a spectator, respectively.

Schematically the total cross section
can be expressed as
\begin{eqnarray}
\sigma^{\rm tot} (N(\ell,\ell'N'))
&=&
\sigma^{\rm sp} (N \rightarrow N') +
\sum_{\Delta} \sigma^{\rm sp} (N \rightarrow \Delta \rightarrow N')
\nonumber \\
&+&
Z \bigl[ \sigma^{\rm h} (N \rightarrow N') +
\sum_{\Delta} \sigma^{\rm h} (N \rightarrow \Delta \rightarrow N') \bigl]
\nonumber \\
&+&
P_{\pi N} \bigl[
\sum_{N"} f_{N"} \sigma^{\rm h} (N" \rightarrow N') +
\sum_{N",\Delta} f_{N"} \sigma^{\rm  h} (N" \rightarrow \Delta \rightarrow
N') \bigr]
\nonumber \\
&+&
P_{\pi \Delta} \bigl[
\sum_{\Delta} f_{\Delta} \sigma^{\rm h} (\Delta \rightarrow N') +
\sum_{\Delta,\Delta'} f_{\Delta} \sigma^{\rm h}
(\Delta \rightarrow \Delta' \rightarrow N)
\bigl] \; .
\label{consum}
\end{eqnarray}
In the formula above $\sigma^{\rm sp}$ and $\sigma^{\rm h}$ are cross
sections for the spectator and colour neutralization mechanisms,
respectively.
The factors $Z=\cos^{2}\theta, P_{\pi N}, P_{\pi \Delta}$
are probabilities of the core, and $\pi N$ and $\pi \Delta$ Fock
components. The $f_{N'}$ and $f_{\Delta}$ factors are the relative weights
of the nucleons (deltas) in the meson cloud model as dictated
by the isospin symmetry. A fully consistent calculation would require to
include the momentum distribution of virtual baryons in the nucleon,
which goes beyond the scope of the present paper.
In the actual calculation we have taken into account only the first two lines
in Eq.(\ref{consum}).
We note that the nucleon can be produced
in the direct way ($M = \pi, B = N$) or as a two step process:
direct production of isobar $\Delta$ ($M = \pi, B =\Delta$)
and subsequent decay of the resonance ($\Delta \rightarrow N + \pi$).

In calculating $\sigma^{\rm h}$ (see Eq.(\ref{consum}))
one should bear in mind that part of the
the sea quark
distributions in bare nucleons (deltas)
is already taken into account by the pionic component.
Therefore  in Eq.(\ref{semiinclcs}) the sea structure functions $S_f(x)$
should be replaced by the difference between the experimental ones
and the ones computed in the pion spectator approach. In practice
we find that at $Q^2= $ 4 (GeV$/c)^2$ this is equivalent with a reduction
of the four quark fragmentation by
a factor $0.5 - 0.7$ \cite{SSG93,HNSS94}.

The probability density to find a meson with longitudinal
momentum fraction $y_{M}$ and four-momentum
of the virtual meson squared, $t$,
(or alternatively transverse momentum $p_{T}^2$), referred to as
splitting function, quantifies the presence of virtual mesons
in the nucleon.
The splitting function $f(y_{M},t)$ for the first process is
\begin{equation}
f_{\pi N} (y_{M},t) = \frac{3 g_{p \pi^0 p}^2}{16\pi^2}
y_{M} \frac{ (-t) |F(y_{M},t)|^2 } {(t - m_{\pi}^2)^2} \label{splitpiN}.
\end{equation}
The splitting function for the $\pi \Delta$ Fock state is
\begin{equation}
f_{\pi \Delta} (y_{M},t) = \frac{2 g_{p \pi^- \Delta^{++}}^2}{16\pi^2}
y_{M} \frac{ (M_{+}^2 - t)^2 (M_{-}^2 - t) |F(y_{M},t)|^2 }
{ 6 m_N^2 m_{\Delta}^2 (t - m_{\pi}^2)^2 } ,  \label{splitpiDelta}
\end{equation}
where $M_{+} = m_{\Delta} + m_N$ and $M_{-} = m_{\Delta} - m_N$ .
The $F(y_{M},t)$ are vertex form factors, which account for the extended
nature of hadrons involved. The form factors used in meson exchange
models are usually taken to be a function of $t$ only.
As discussed in Refs.\cite{Z92,HSS93}
such form factors are a source of momentum sum rule violation.

The pion (meson) emitted by the nucleon can interact with a virtual
$W$-boson. In general the cross section for the semi-inclusive   spectator
$N(\ell,\ell'B)X$ process depicted in Fig.1 can be written as
\begin{equation}
\frac{d \sigma^{\rm sp} \bigl( N(\ell,\ell'B)X \bigr) } {dx dQ^2 dy_{M}
dp_{T}^2}
= f_{M/N}(y_{M},p_{T}^2) \frac{d \sigma^{\ell M} } {dx dQ^2} (x/y_{M})
\; .
\end{equation}
Integrating over unmeasured quantities one recovers the known
one-pion exchange contribution to the inclusive cross section
\begin{equation}
\frac{ d\sigma^{\ell N}(x) } {dx} =
\int_{x}^{1} dy_{M} f_{\pi}(y_{M})
\frac{d\sigma^{\ell \pi}(x/y_{M})} {dx} ,
\end{equation}
where the $\frac{d\sigma^{\ell \pi}}{dx}$ is the $Q^2$-integrated
cross section for the deep inelastic scattering of $\ell$ from
the virtual pion. In practical calculations the on-mass-shell $\ell \pi$
cross section given by Eq.(\ref{in_on_shell}) can be used.

The probability to find a nucleon $N'$ associated with the lepton
$\ell'$ at given value of the Bjorken $x$ can easily by calculated as
\begin{eqnarray}
P_{N'}^{\ell N}(x) =
|< 1 t_{3}^{\pi}, {1 \over 2} t_{3}^{N'} | {1 \over 2} t_{3}^{N} >|^2
\int_x^1 \, dy_M
f_{\pi N'/N}(y_{M}) {\cal G}_{\rm cc}^{\ell \pi}(x/y_{M})  \nonumber \\
 +
\sum_{t_{3}^{\Delta}}
|< 1 t_{3}^{\pi}, {3 \over 2} t_{3}^{\Delta} | {1 \over 2}t_{3}^{N}>|^2
\int_x^1 \, dy_M f_{\pi \Delta/N}(y_{M}) {\cal G}_{\rm cc}^{\ell \pi}(x/y_{M})
P(\Delta \to \pi N').
\label{semi-i_x}
\end{eqnarray}
The first term in Eq.(\ref{semi-i_x}) describes the direct nucleon
production whereas the second term corresponds to the two step mechanism.
The Clebsch-Gordan coefficients account for the relative yield
of a given isospin channel.
The sum involves all charge combinations of the pion and $\Delta$ which
lead to the final nucleon $N'$.
The probability of the $\Delta$ resonance decay into the channel of interest
($N'$) can be calculated assuming isospin symmetry
\begin{equation}
P(\Delta \to \pi N') \; = \;
|< 1 t_{3}^{\pi}, {1 \over 2} t_{3}^{N'} | {3 \over 2} t_{3}^{\Delta} >|^2 .
\label{semi_i_x}
\end{equation}

The formula given by Eq.(\ref{semi-i_x}) can be easily generalized
to the case with the proton (nucleon) in a given momentum range
($p_{\rm min} < p < p_{\rm max}$).
In this case the following substitutions are necessary
\begin{eqnarray}
f_{\pi N'/N}(y_{M})
& \to &
\int_{t_{\rm min}^N}^{t_{\rm max}^N}
f_{\pi N'/N}(t,y_{M}) \, dt \; , \nonumber \\
f_{\pi \Delta /N}(y_{M})
& \to &
\int_{t_{\rm min}^{\Delta}}^{t_{\rm max}^{\Delta}}
f_{\pi\Delta/N}(t,y_{M}) P(t,p_{\rm min},p_{\rm max}) \, dt \; .
\end{eqnarray}
For the direct process the experimental limits on the proton momentum
directly correspond to limits for t:
\begin{eqnarray}
t_{\rm min}^{N} &=& 2[m_N^2 - m_N \sqrt{ p_{\rm max}^2 + m_N^2 }],
\nonumber \\
t_{\rm max}^{N} &=& 2[m_N^2 - m_N \sqrt{ p_{\rm min}^2 + m_N^2 }],
\end{eqnarray}
with an extra absolute limit $t < -m_N^2 y_{M}^2 / (1-y_{M})$.
For the proton production through the intermediate delta resonance one
has to integrate over all possible $t$ and include an extra probability
function $P(t,p_{\rm min},p_{\rm max})$.
The $P(t,p_{\rm min},p_{\rm max})$ is the probability that the intermediate
isobar $\Delta$ produced with momentum
\begin{equation}
p_{\Delta}^2(t) =
\left[ \frac{m_{\Delta}^2 + m_N^2 - t} { 2 m_N } \right]^{2} -
                  m_{\Delta}^2
\end{equation}
decays into a pion and nucleon $N'$ in the given momentum range.
It can easily be calculated assuming a sharp $\Delta$ resonance and its
isotropic decay.  Technically it is convenient to calculate
$P(t,p_{\rm min},p_{\rm max})$ as a difference
\begin{equation}
P(t,p_{\rm min},p_{\rm max}) = P(t, p_{N'} < p_{\rm max}) - P(t, p_{N'}
    < p_{\rm min})
\end{equation}
with self-explanatory notation.
In  this approximation the probability is
\begin{equation}
P(t, p_{N'} < p_{\rm max}) = \frac{1}{2} \Bigl[ 1 -
\frac{ E_{0} E_{\Delta}/m_{\Delta} - (p_{\rm max}^2 + m_{N}^2)^{1/2} }
    {p_{0} \bigl(E_{\Delta}/m_{\Delta} - 1 \bigr)^{1/2} } \Bigr] \; ,
\end{equation}
where $p_{0}$, $E_{0}$ are momentum and energy of the nucleon
in the delta rest frame.

The momentum distribution of the spectator baryon $B$ produced in
the $N(\ell,\ell'B)X$ reaction which proceeds via scattering
of virtual boson ($\gamma, W_{\pm}$) on the virtual meson $M$ can
generally be written as
\begin{equation}
\frac{dN(p,Q^{2})}{dp} =
\frac {
 \frac{1}{\Delta p}
\int dx \int_{x}^{1} dy_{M}
\big[
\int_{t_1}^{t_2} f_{MB/N}(y_{M},t) dt
\big]
\;\;
\frac{ d\sigma_{M}^{\ell \ell'} } {dx} ({x \over y_{M}}, Q^2)
}
{ \int dx \frac{d\sigma_{N}^{\ell \ell'} } {dx} (x,Q^2) }
\label{mom_dis}
\end{equation}
with $t_1$ and
$t_2 $ corresponding to momenta $p \pm \Delta p/2 ,$ respectively.
The function $f_{MB/N}(y_{M},t)$ can be calculated for different
partitions of the nucleon into a meson $M$ (pseudoscalar,vector)
and baryon $B$ (octet,decuplet) \cite{HSS93}.

\newpage


\section{Results and Discussion}


Let us first discuss  the probability for production  of slow protons as
obtained from the hadronization models. (With slow protons we  mean
hereafter unless specified otherwise, protons with momenta in
the interval (0.15 - 0.60 GeV$/c$), corresponding to
the experimental bins defined in Ref.\cite{G89}).
In Table 2  we list the probabilities  calculated in
(a)  the quark-diquark model from Ref.\cite{BDT93},
(b)  the quark-diquark model corrected for the presence of
    the pionic cloud (to be discussed later) and
(c) in the Lund string model \cite{I92}.
In the first two cases  the results are rather sensitive to the form
of the fragmentation function $D(\tilde{z})$. In this calculation
the transverse momentum distribution function $\Phi (p_T^2)$ in
Eq.(\ref{phi}) is parametrized by a Gaussian form
\begin{equation}
\Phi(p_T^2) = \frac{1}{\sigma ^2} \exp(-p_T^2 / \sigma^2)
\end{equation}
with $\sigma ^2$ = 0.3 (GeV$/c)^2$.
We find that the use of a constant fragmentation function
 (as in Ref.\cite{MTN92})
yields a result almost consistent
with the  experimental one \cite{G89}. The rate obtained with
quadratic $D(\tilde{z}) = 1/2*\tilde{z}(1-\tilde{z})$ and especially
with the `triangular'
$D(\tilde{z}) = 2(1-\tilde{z})$ (counting rule motivated for the diquark
hadronization)
hadronization functions is lower than
the experimental rate. The quark-diquark model \cite{BDT93} gives only
isospin factors for the valence quark (diquark) hadronization.

To show the sensitivity to details of the sea (4-quark)
fragmentation two extreme limits have been used in which protons
being produced either with zero probability
(only diquark hadronization) or with unit probability
(neglecting production of other baryons),
respectively.
As is seen from the results the contribution of the sea hadronization
is rather
important for the production of slow protons. The total rate obtained
in both extreme limits is, however, below the experimental rate
\cite{G89}.
The Lund string model predicts an even lower rate.

This supports the conjecture  that another mechanism plays a 
role and therefore
we  now turn to a discussion of the role of pion exchange
mechanism as sketched in the previous section.

In all calculations discussed below  exponential vertex form factors
\begin{equation}
F(y_{M},p_{T}^2) =
\exp[- \frac{M_{MB}^2(y_{M},p_{T}^2) - m_N^2} {2\Lambda^2} ]
\label{formfac}
\end{equation}
have been used in Eqs.(\ref{splitpiN},\ref{splitpiDelta}).
In Eq.(\ref{formfac}) $M_{MB}(y,p_{T}^2)$ is
the invariant mass of the intermediate meson-baryon system.
The cut-off parameters used in the present calculation
($\Lambda_{\pi N}$ = 1.10 GeV and $\Lambda_{\pi \Delta}$ = 0.98 GeV)
have been determined from the analysis of the particle spectra
for high-energy neutron ($p(p,n)X$) and $\Delta$
( $p(p,\Delta^{++})X$ ) production \cite{HSS94}.
With these cut-off parameters
the NMC result for the Gottfried sum rule \cite{A91}
which depends sensitively on $\Lambda$, has been
reproduced \cite{SH93,HSS94}. Furthermore the
model describes $\overline{u}/\overline{d}$ asymmetry extracted
recently from the Drell-Yan NA51 experiment at CERN \cite{HNSS94}.
Finally the deep-inelastic structure functions of pions are taken from
Ref.\cite{B83} and those for the nucleon from Ref.\cite{O91}.

A quantity of interest \cite{BDT93} is the probability for finding a
slow proton in a momentum bin $\Delta p$ as a function of $x$
\begin{equation}
P(x,\Delta p) = \frac
{d \sigma(x, \Delta p)}{dx}  (\frac{ d \sigma}{dx})^{-1} .
\label{P(x)}
\end{equation}
The quantity calculated in the spectator model is shown in
Fig.2 where in addition we show the direct component by the dashed
line (the contribution of the intermediate $\Delta$ is a complement to
the total $P(x)$ given by the solid line) and the contribution from
the valence quarks in the pion by the dotted line (the sea
contribution is again a complement to the solid line). As seen from
the figure the direct component dominates the cross section in
the whole range of $x$. The contribution of the sea in the pion plays
an important role only at small Bjorken $x$. Experimentally this
region can only be accessible at high beam energies.

In order to obtain an absolutely normalized "experimental" quantity
the following procedure has been applied.
In Ref.\cite{G89} the ratios of normalized $x$-distributions with
and without secondary protons in the momentum range $\Delta p$
\begin{equation}
R(x,\Delta p) = \frac
{ \frac{d \sigma} {dx} (x,\Delta p) / \sigma(\Delta p) }
{ \frac{d \sigma} {dx} (x,1- \Delta p) / \sigma(1 - \Delta p) }
\end{equation}
have been given, where
 $(1 - \Delta p)$ refers to the complement of the momentum interval
$\Delta p$. Taken the absolute normalization
$R_{\rm tot} = \sigma (\Delta p)/ \sigma_{\rm tot}$ from Table 1 of
Ref.\cite{G89} at face value one can extract the absolutely normalized
 quantity (\ref{P(x)}) as
\begin{equation}
P(x,\Delta p) = \frac
{R(x,\Delta p)} {R(x,\Delta p) + (1 - R_{\rm tot})/R_{\rm tot} }.
\end{equation}
The results of the procedure are presented in Fig.3 for
low (0.15 -0.35 GeV/c) and high (0.35 -0.60 GeV/c) momentum bins.
The results of
the spectator model (dashed line) and of
the Lund string model\cite{I92} (solid line) are shown in Fig.3.
As can be seen from the figure the string breaking mechanism
underestimates the experimental data.
To get more insight into this result it is instructive to consider
simpler quark-diquark model \cite{BDT93} for the flavour dependent
diquark branching ratio into nucleons and deltas. In this calculation
we have taken the QCD motivated fragmentation functions \cite{FS81}
\begin{equation}
D_{2q}(\tilde{z}) = C \, \tilde{z}^{1/2}(1-\tilde{z}), \; D_{4q}(\tilde{z}) =
  \mbox{ const}.
\label{QCD_mot}
\end{equation}
Since no realistic model for the four-quark branching ratio exists,
one can only consider lower (diquark fragmentation) and
upper (maximal proton production in four-quark hadronization) estimates,
as for the relative yields in Table 2.
The results are shown in Fig.3 as dotted lines. The difference
between the lower and the upper limit clearly demonstrates a significant
sensitivity to the four-quark hadronization. The presence of
the virtual pions leads to a reduction of the perturbative sea by about
a factor 2 \cite{HNSS94}, which would result in the same reduction
of the four-quark contribution to $P(x).$
Even the upper limit underestimates the experimental data.
This suggests a presence of an extra competing mechanism.

For the proton production on the neutron the situation is somewhat
more complicated (see Fig.4). Here the result for the Lund string model
falls only slightly below the experimental data \cite{BDT93}
extracted from the $(\nu,\mu^{-})$ reaction on the deuteron (Fig.4a,b).
The spectator mechanism predicts a very similar result, also almost
consistent with the experimental data.
For the $(\overline{\nu},\mu^{+})$
reaction (Fig.4c,d) the situation is similar, except that here the
corresponding cross sections are significantly smaller.
Similarly as for the proton production on the proton we present
the lower (diquark fragmentation) and
the upper (maximal proton production in four-quark hadronization)
limits of the model discuss in section 2.2.
Again a large difference between the two limits can be observed.
The proton production on the neutron , however,
differs from that on the proton.
It is reasonable to expect here rather the lower limit to be more
realistic. The experimental data seems to support this
expectation.

In order to better understand the range of  proton momenta where
the spectator mechanism plays an important role, it is instructive
to calculate the momentum distribution of protons produced in
the neutrino (anti-neutrino) induced reactions.
The proton momentum distributions were determined
in the early BEBC experiments at CERN \cite{A83}. In order to avoid
the resonance contributions and minimize the effect of the proton's sea,
the events have been selected with the constraints $W >$ 3 GeV and
$x >0.1$.

The cuts are rather important especially for the low-momentum
part of the spectrum. In the experiment \cite{A83} the wide band
(anti-)neutrino beam has been used. In principle, the effect of
the beam spread should be taken into account. It has been checked,
however, that the results of the Lund string model\cite{I92} are
not very sensitive to the beam energy.  The results of
the Monte Carlo simulation for three different beam energies
(20, 50, 80 GeV) are shown in Fig.5.
The total number of protons per event in the momentum
range (0.15 $ < p <$ 0.60 GeV$/c$) and
corresponding to the experimental cuts\cite{A83}, calculated in
the Lund string model and the spectator model is given in Table 3.
Also shown are the experimental results extracted from Ref.\cite{A83}.
Similarly as for the experiment from Ref.\cite{G89}
the string mechanism produces too little low energy protons.
In this case the spectator mechanism gives almost total missing
strength (see Table 3).

In order to get insight into the total proton momentum distribution,
it is instructive to consider the simpler quark-diquark model \cite{BDT93}
for the flavour dependent diquark branching ratio into nucleons and
deltas. The momentum distribution calculated with the QCD motivated
fragmentation functions (Eq.(\ref{QCD_mot}))
is shown in Fig.6. It was assumed here for simplicity that the
four-quark hadronization leads in all cases to  proton
production. This will lead to an overestimation of this contribution.
The diquark and four-quark hadronization contributions
are shown separately by the dashed and dotted line, respectively.
As clearly seen from the figure, the four-quark hadronization
populates the slow proton region, whereas the diquark hadronization
is responsible for the production of rather fast protons.

The momentum distribution of protons produced in the spectator
mechanism calculated according to Eq.(\ref{mom_dis}) is shown
in Fig.7. The solid line includes both protons
originating from the $\pi^0 p$ Fock component (direct component)
and those originating from the decay of spectator deltas.
The direct component is shown separately by the dashed line.
In the neutrino induced reaction the sequential mechanism is as
important as the direct production of protons, whereas in antineutrino
induced reaction its relative contribution is much smaller.
This effect is a consequence of the dependence of underlying
coupling constants on the charge of pions and of the difference of
$F_2^{\nu,\pi}(x)$ and $F_2^{\overline{\nu},\pi}(x)$ structure
functions. In order to better understand the underlying reaction
mechanism we present also in Fig.7 separately the contribution from
the valence quarks in the pion (dotted line). The corresponding
contribution from the sea is a complement to the total momentum
distribution.

In the experiment of ref.\cite{A83} not
all protons with low momenta ( $p <$ 0.6 GeV$/c$)
were identified.
In order to make comparison of the model results in the whole momentum
range the inclusion of the proton identification efficiency is
neccessary. In the present paper the proton identification
efficiency has been parametrized as
\begin{equation}
e(p) = \Bigl[ 1 + \exp[ \frac{p-p_0} {\sigma} ] \Bigr]^{-n}
\end{equation}
with $p_0$ = 0.86 GeV/$c$, $\sigma$ = 0.1 GeV/$c$ and $n$ = 0.25.
The result of the spectator mechanism and the result of the
quark-diquark model have been multiplied by the efficiency function
$e(p)$ and are shown in Fig.8 by the dashed and dotted line,
respectively. One clearly sees that the colour neutralization mechanism
(dotted line) is dominant for the proton production in a broad
momentum range. The spectator mechanism (dashed line) plays, however,
important role at very low momenta.
Since in our model the pionic cloud constitutes about half
of the nucleon sea \cite{HNSS94} for $Q^2$ of a few GeV$^2$,
for consistency the four-quark hadronization contribution
(dotted line in Fig.6) must be reduced by a factor of about 0.5.
Therefore the sum
\begin{equation}
N(p) = e(p) \; [N_{2q}(p) + 0.5 N_{4q}(p) + N_{\pi}(p)]
\end{equation}
consistently includes the two mechanisms.
The sum (solid line in Fig.8) almost coincides with the original
quark-diquark model and describes the experimental data \cite{A83}
in a resonable way.

The considered $x$-distributions $P(x)$ and proton momentum distributions
do not give a direct evidence for identifying the meson cloud in
the nucleon, although they do not contradict its existence.
Therefore one should look for other possibilities. It has been suggested
by some authors to consider the production of slow (for the fixed target
experiments) pions. Slow pions could be created by a mechanism analogous
to that considered in the present paper for the slow proton production,
as a spectator of the reaction (see Fig.9). The momentum distribution
of pions can be calculated in a fully analogous way to that of
the slow protons (see formula (\ref{mom_dis}) ).  In this case
all kinematical variables
associated with the meson $M$ (pion) have to be replaced by
corresponding variables for the recoil baryon $B$ (nucleon or delta).
The momentum distributions of pions ($\pi^{+}, \pi^{0}, \pi^{-}$)
produced in the $(\nu, \mu^{-})$ and $(\overline{\nu}, \mu^{+})$
reactions calculated in this way are shown in Fig.10, separately for
$N \pi$ (solid line) and $\Delta \pi$ (dashed line) mechanisms.
For comparison the momentum distribution of pions generated by the Monte
Carlo method \cite{I92} according to the the Lund string model is shown
by the solid line with the centered points. As seen from the figure the
spectator pions constitute only a small fraction of all pions, therefore
their identification seems to be rather difficult.

The relative fraction of different pions differs between the Lund
string model \cite{I92} and the spectator model (see Table 4).
This is especially visible for the neutrino induced reaction on
the proton and antineutrino induced reaction on the neutron.
The spectator mechanism predicts a huge asymmetry between $\pi^{+}$
and $\pi^{-}$ production. The reason here is very simple.
Both $\pi^{-}$ in the $\nu + p$ and $\pi^{+}$ in
the $\overline{\nu} + n$ reactions
cannot be produced in the scattering of the virtual boson
$W^{\pm}$ off the nucleon of the $\pi N$ Fock states. On
the other hand their production via scattering of $W^{\pm}$ from the
virtual delta of the $\pi \Delta$ Fock state is highly suppressed as
here the corresponding structure functions vanish
in the valence quark approximation.  The huge difference between
the predictions of the Lund string model and the spectator mechanism
could in principle be tested if one limits the analysis only to the
pions produced in the backward directions. This part of the
phase space is expected to be dominated via the spectator mechanism.
A possible confirmation of the asymmetry in the production of the
positively and negatively charged pions would be a strong test of the
spectator mechanism and the concept of virtual pions in the nucleon.

\newpage


\section{Conclusions}

In the present paper the production of slow protons produced
in (anti)neutrino induced reactions on proton and neutron has been
studied. We have analyzed the total rate, the Bjorken-$x$
dependence of the slow proton production as well as
the momentum distribution of emitted protons.

The  analysis of the rate of slow proton production
in the neutrino and anti-neutrino charged current reactions
shows that "realistic" hadronization models underestimate
the cross section measured by the bubble chamber Collaboration at
CERN \cite{G89}.
Although the pion exchange mechanism leads  to a rather small contribution
to the total proton production, it plays an important role in
the production of slow protons and helps us to understand the deficit
of the slow proton production as predicted by standard hadronization
models based on colour neutralization mechanism.

Even these two mechanisms seem to fail to describe the total
rate of slow proton production from Ref.\cite{G89}.
Comparison of the total rate as extracted here from earlier
CERN measurement \cite{A83}, where more severe restrictions
on the Bjorken-$x$ and invariant mass of the produced hadrons
have been imposed, suggests that the `missing' mechanism is
related to the region of small $x$ and low invariant mass of the produced
hadrons.
The considered $x$-distributions $P(x)$ and proton momentum distributions
for (anti-)neutrino charged current reactions do not give  direct
evidence for identifying the meson cloud in the nucleon,
although they do not contradict its existence.
Therefore one should look for other possibilities to prove or disprove
the whole concept of the meson cloud in the nucleon.

The recent results of deep inelastic electron-proton scattering
at HERA, (ref\cite{HERA}), open such a possibility. The observation of a
surprisingly large number of events
characterized by a large rapidity gap, strongly
suggests the presence of a (up to now not really identified) proton
with approximately beam velocity. These so-called diffractive scattering
events, which cannot be explained by conventional hadronization codes,
have up to now primarily been interpreted in terms of Pomeron exchange.
Since the Pomeron by definition has vacuum quantum numbers this approach
predicts that only protons are produced in the final state
in contrast to the mesonic cloud
model which leads to a definite prediction for branchings into proton,
neutron and isobars. Therefore the planned search for neutron spectators
will provide more information on the mechanism for the production
of slow baryons.

$\it {Acknowledgements.}$ \\
 Stimulating discussions with T.Coghen, J.Guy, A.G.Tenner and
W.Wittek are gratefully acknowledged.
Special thanks are due to K.Golec for providing us with the code
LEPTO 6.1.
This work was supported
by the The Netherlands Foundation for Fundamental Research (FOM),
and  by the Polish KBN grant 2 2409 9102.



\newpage


\setlength{\parindent}{0.0cm}
\begin{center}
{\bf Tables}
\end{center}

\begin{center}

Table 1. Fraction  of protons ($f_p = N_p/(N_p+N_n)$) and neutrons
($f_n = N_n/(N_p+N_n)$) produced in the $e+p$, $\nu+p$ and
$\overline{\nu} + p$ reactions. The fractions have been calculated
according to the quark-diquark model \cite{BDT93} and the Lund string
model \cite{I92} for the beam energy 50 GeV. The total number
of protons and neutrons per event in the Lund string model exceeds
unity which is caused by the explicit production of
baryon-antibaryon pairs.

\vskip 0.5cm

\begin{tabular}{|l|c|c|c|c|}
 \hline
   beam & $p \rightarrow p$ & $p \rightarrow n$
          & $n \rightarrow p$ & $n \rightarrow n$ \\
 \hline
 quark-diquark model   &        &        &        &   \\
 $ \mu^{\pm}   $       & 0.5473 & 0.4527 & 0.2160 & 0.7840 \\
 $ \nu   $             & 0.9259 & 0.0741 & 0.5000 & 0.5000 \\
 $ \overline{\nu} $    & 0.5000 & 0.5000 & 0.0741 & 0.9259 \\
 \hline
 Lund string model     &        &        &        &        \\
 $ \mu^{\pm}   $       &  0.60  &  0.47  &  0.43  &  0.64  \\
 $ \nu  $              &  0.78  &  0.29  &  0.56  &  0.52  \\
 $ \overline{\nu} $    &  0.52  &  0.54  &  0.31  &  0.76  \\
                       &        &        &        &        \\
 \hline
\end{tabular}

\vskip 2cm

\newpage

Table 2. Fraction (in percent) of events with low (0.15  $< p <$ 0.60 GeV/c)
momentum protons for neutral current charged lepton DIS
and charged current neutrino and antineutrino reactions.
\\

\vskip 0.5cm

\begin{tabular}{|l|c|c|c|c|}
 \hline
 final state baryons & $D(z)$ & $ p\;+\; \mu^{\pm} \to p$ &
 $ p\;+\;\nu \to p$ & $ p\;+\;\overline{\nu} \to p$ \\
 \hline
%
 simple    quark-diquark        &            &       &       &       \\
 model     & constant   & 3.47-5.88 & 6.70-9.75 & 2.74-6.57 \\
           & quadratic  & 2.15-3.89 & 4.22-6.42 & 1.69-4.45 \\
           & triangular & 0.89-1.64 & 1.76-2.70 & 0.70-1.89 \\
 \hline
 improved quark-diquark         &	     &	     &	     &	     \\
 model           & constant   & 3.43-5.54 & 5.72-8.04 & 2.75-6.13 \\
           & quadratic  & 2.13-3.65 & 3.59-5.26 & 1.70-4.14 \\
           & triangular & 0.88-1.54 & 1.50-2.21 & 0.70-1.75 \\
 \hline
  Lund string model         &            &       &       &       \\
   $E$ = 20 GeV                  &            &  2.97 &  0.89 &  1.44 \\
  $ E$ = 50 GeV                  &            &  4.37 &  1.43 &  1.95 \\
  $ E$ = 80 GeV                  &            &  4.60 &  1.42 &  1.75 \\
 \hline
  spectator model              &            &       &       &       \\
 $ \pi N $                     &            &  0.97 &  1.23 &  1.55 \\
 $ \pi \Delta $                &            &  0.29 &  0.52 &  0.27 \\
                               &            &       &       &       \\
  valence                      &            &  0.88 &  1.33 &  1.28 \\
  sea                          &            &  0.38 &  0.42 &  0.53 \\
                               &            &       &       &       \\
  sum                          &            &  1.25 &  1.75 &  1.81 \\
 \hline
 Exp $\cite{G89} $            &            &       &  10   &    8  \\
 \hline
\end{tabular}

\newpage

Table 3. Fraction (in percent) of events with protons in the momentum interval
0.15 $< p <$ 0.60 GeV$/c$ for the neutrino and antineutrino
induced reactions. The results from the Lund string model
has been obtained with extra limitations: $x > 0.1$ and W $>$ 3 GeV
for three different beam energies.
\\
\vskip 0.5cm

\begin{tabular}{|l|c|c|}
 \hline
    contribution  & $ p\;+\;\nu \to p$ & $ p\;+\;\overline{\nu} \to p$ \\
 \hline
 spectator mech.       &        &          \\
 $ \pi N $             &  1.23  &  1.55    \\
 $ \pi \Delta $        &  0.52  &  0.27    \\
  sum                  &  1.75  &  1.81    \\
 \hline
 Lund string model     &        &          \\
   E = 20 GeV          &  0.98  &  1.39    \\
   E = 50 GeV          &  1.27  &  1.79    \\
   E = 80 GeV          &  1.17  &  1.58    \\
 \hline
  Exp.$\cite{A83} $    &  4.74  &  5.20    \\
 \hline
\end{tabular}

\vskip 2cm

Table 4.
Fractions of different pions calculated in the Lund string
model \cite{I92} for the beam energy 50 GeV and in the spectator
mechanism.

\vskip 0.5cm

\begin{tabular}{|l|c|c|c|c|c|c|}
\hline
& \multicolumn{3}{c|}{Lund string model}
& \multicolumn{3}{c|}{spectator mechanism} \\
 \cline{2-7}
 reaction & $\pi^{-}$ & $\pi^{0}$ & $\pi^{+}$
          & $\pi^{-}$ & $\pi^{0}$ & $\pi^{+}$
\\
 \hline
 $ \mu^{\pm} + p$
  &  0.29  &  0.36  &  0.35  &  0.25  &  0.38  &  0.37  \\
 $ \nu + p$
  &  0.23  &  0.36  &  0.41  &  0.01  &  0.24  &  0.75  \\
 $ \overline{\nu} + p$
  &  0.36  &  0.37  &  0.27  &  0.29  &  0.40  &  0.31  \\
 \hline
 $ \mu^{\pm} +n$
  &  0.36  &  0.36  &  0.28  &  0.63  &  0.28  &  0.09  \\
 $ \nu +n$
  &  0.28  &  0.36  &  0.35  &  0.29  &  0.41  &  0.31  \\
 $ \overline{\nu} +n$
  &  0.42  &  0.36  &  0.22  &  0.71  &  0.25  &  0.04  \\
 \hline
\end{tabular}

\pagebreak


\newpage
{\large\bf {Figure Captions}}\\
\begin{description}

\item[Fig.1]
The spectator mechanism for the production of slow protons in
the (anti)neutrino induced reactions on the nucleon.

\item[Fig.2]
$P(x,\Delta p)$ for 0.15 $< p <$ 0.60 GeV/c
calculated within the spectator model (solid line).
The contribution of the direct proton production is shown
by the dashed line and the contribution from the valence quarks in
the pion by the dotted line.

\item[Fig.3]
$P(x,\Delta p)$ for $p(\nu,p)X$ and $p(\overline{\nu},p)X$
reactions, for low
(0.15 $< p <$ 0.35 GeV/c) and high
(0.35 $< p <$ 0.60 GeV/c) momentum bins.
The "experimental" data has been obtained by the procedure
described in the text.
The result from the spectator model is shown by
the solid and from the Lund string model by the dashed line.
In addition the two limits discussed in the text are shown
by the dotted lines.

\item[Fig.4]
$P(x,\Delta p)$ for $n(\nu,p)X$ and $n(\overline{\nu},p)X$ for low
(0.15  $< p <$ 0.35 GeV/c) and high
(0.35  $< p <$ 0.60 GeV/c) momentum bins.
The experimental data are taken from Ref.\cite{BDT93}.
The result from the spectator model is shown by
the solid and from the Lund string model by the dashed line.
In addition the two limits discussed in the text are shown
by the dotted lines.

\item[Fig.5]
Momentum distribution of protons from the Lund string model
for different (anti)neutrino beam energies
($E_{\rm beam}$ = 20 GeV -- dashed line,
 $E_{\rm beam}$ = 50 GeV -- solid line and
 $E_{\rm beam}$ = 80 GeV -- dotted line.
The experimental \cite{A83} cuts: $W > 3 $GeV and $x > 0.1$ have been
taken into account. The proton identification efficiency of
the experiment is not included here.

\item[Fig.6]
Momentum distribution of protons from the quark-diquark model
\cite{BDT93}. The diquark fragmentation contribution is shown
by the dashed line and the 4-quark fragmentation contribution
by the dotted one. The details concerning the fragmentation function
used are described in the text.

\item[Fig.7]
Momentum distributions of protons produced in the spectator mechanism
in the neutrino (a) and antineutrino (b) induced deep inelastic
scattering (solid line). The direct $\pi N$ contribution is shown
separately by the dashed line, whereas
the contribution from the valence quarks
in the pion is shown by the dotted line.

\item[Fig.8]
Momentum distributions of protons corrected for the identification
efficiency from the quark-diquark model (dotted line)
and from the spectator model (dashed line)
compared with the CERN experimental data \cite{A83}.
The solid line includes both mechanisms in the way
described in the text.

\item[Fig.9]
The spectator mechanism for the production of slow pions
in the (anti)neutrino induced reactions on the nucleon.

\item[Fig.10]
Momentum distribution of pions produced in the $p(\nu,\pi)X$ and
$p(\overline{\nu},\pi)X$ deep-inelastic processes.
The predictions of the spectator model are shown as the solid ($\pi N$)
and the dashed ($\pi \Delta$) lines. The result of the Monte Carlo
simulation according to the Lund string model \cite{I92} are shown
by the solid line with the centered points.

\end{description}
\end{center}

\end{document}